\magnification=\magstep1 
\font\bigbfont=cmbx10 scaled\magstep1
\font\bigifont=cmti10 scaled\magstep1
\font\bigrfont=cmr10 scaled\magstep1
\vsize = 23.5 truecm
\hsize = 15.5 truecm
\hoffset = .2truein
\baselineskip = 14 truept
\overfullrule = 0pt
\parskip = 3 truept
\def\frac#1#2{{#1\over#2}}

\nopagenumbers

\topinsert
\vskip 3.2 truecm
\endinsert
\centerline{\bigbfont  MANY-BODY EFFECTIVE MASS ENHANCEMENT}
\vskip 6 truept
\centerline{\bigbfont IN A TWO-DIMENSIONAL ELECTRON LIQUID}
\vskip 18 truept
\centerline{\bigifont R. Asgari}
\vskip 8 truept
\centerline{\bigrfont NEST-INFM and Classe di Scienze, Scuola Normale Superiore}
\vskip 2 truept
\centerline{\bigrfont I-56126 Pisa, Italy}
\vskip 2 truept
\centerline{\bigrfont Institute for Studies in Theoretical Physics and Mathematics}
\vskip 2 truept
\centerline{\bigrfont Tehran 19395-5531, Iran}
\vskip 14 truept
\centerline{\bigifont B. Davoudi}
\vskip 8 truept
\centerline{\bigrfont D\'epartement de Physique, Universit\'e de Sherbrooke}
\vskip 2 truept
\centerline{\bigrfont Sherbrooke, Qu\'ebec, Canada J1K 2R1}
\vskip 2 truept
\centerline{\bigrfont Institute for Studies in Theoretical Physics and Mathematics}
\vskip 2 truept
\centerline{\bigrfont Tehran 19395-5531, Iran}
\vskip 14 truept
\centerline{\bigifont M. Polini and M. P. Tosi}
\vskip 8 truept
\centerline{\bigrfont NEST-INFM and Classe di Scienze, Scuola Normale Superiore}
\vskip 2 truept
\centerline{\bigrfont I-56126 Pisa, Italy}
\vskip 14 truept
\centerline{\bigifont G. F. Giuliani}
\vskip 8 truept
\centerline{\bigrfont Physics Department, Purdue University}
\vskip 2 truept
\centerline{\bigrfont West Lafayette, Indiana 47907, USA}
\vskip 14 truept
\centerline{\bigifont G. Vignale}
\vskip 8 truept
\centerline{\bigrfont Department of Physics and Astronomy, University of Missouri-Columbia}
\vskip 2 truept
\centerline{\bigrfont Columbia, Missouri 65211, USA} 

\input amssym.tex
\vskip 1.8 truecm

\centerline{\bf 1.  INTRODUCTION}
\vskip 12 truept
An electron liquid on a uniform neutralizing background (EL)
is used as the reference system in most realistic calculations of
electronic structure in condensed-matter physics~[1].
At zero temperature there are only two relevant parameters for a disorder-free EL 
in the absence of quantizing magnetic fields and spin-orbital coupling: (i) the usual Wigner-Seitz density parameter $r_s=(\pi n_{\rm \scriptscriptstyle 2D} a^2_B)^{-1/2}$, $a_B$ being the Bohr radius in vacuum (in a medium $a_B\rightarrow a^\star_B=\hbar^2{\bar \kappa}/(m_{\rm b} e^2)$ with ${\bar \kappa}$ and $m_{\rm b}$ appropriate dielectric constant and bare band mass respectively); and (ii) the degree of spin polarization $\zeta=|n_\uparrow-n_\downarrow|/n_{\rm \scriptscriptstyle 2D}$. Here $n_\sigma$ is the average density of particles with spin $\sigma=\uparrow,\downarrow$ and $n_{\rm \scriptscriptstyle 2D}=n_\uparrow+n_\downarrow$ is the total average density.

Understanding the many-body aspects of this model
has attracted continued interest for many decades~[2-6]. 
The EL, unlike systems of classical particles, behaves like an ideal paramagnetic gas 
at high density ($r_s\ll 1$) and like a solid at low density~[7] ($r_s\gg 1$). In the intermediate
density regime, which is relevant in three dimensions ($3D$) to conduction electrons in simple metals and in two dimensions ($2D$) to electrons in an inversion layer of a Si metal-oxide-semiconductor field-effect transistor (MOSFET) or in an AlGaAs/GaAs quantum well, perturbative techniques are clearly not effective owing to the lack of a small expansion parameter. One has to take recourse to approximate semi-analytical methods, a number of which have been reviewed in Refs.~[3,4], or to Quantum Monte Carlo (QMC) simulation methods~[8-18].

Among the methods designed to deal with the intermediate density regime, of particular interest for its physical appeal and elegance is Landau's phenomenological theory~[19]. The basic idea of Landau's theory is that the low-lying excitations of a system of interacting Fermions with repulsive interactions can be constructed starting from the low-lying states of a noninteractig Fermi gas by adiabatically switching-on the interaction between particles. This procedure allows to  establish a one-to-one correspondence between the eigenstates of the ideal system and the approximate eigenstates of the interacting one. Landau called such single-particle excitations of an interacting Fermi-liquid ``quasiparticles" (QP's). He wrote the excitation energy of the Fermi-liquid $E[n_{\bf p}]$ as a functional of the QP distribution function $n_{\bf p}$ in terms of the isolated quasiparticle energy ${\cal E}_{\bf p}$ and of the QP-QP interaction function $f_{{\bf p},{\bf p}'}$. The latter can in turn be used to obtain various physical properties of the system, such as the compressibility and the spin-susceptibility.

One of the implications of Landau's theory of Fermi-liquids is the fact that the QP mass $m^\star$ is renormalized by electron-electron interactions~[2]: $m^\star \neq m$, $m$ being the bare electron mass. In a translationally-invariant system the current ${\bf j}_{\bf p}$ carried by a single excited QP of momentum ${\bf p}$ is controlled only by the bare mass, ${\bf j}_{\bf p}={\bf p}/m$. On the other hand, the QP group velocity ${\bf v}_{\bf p}$ is instead defined by ${\bf v}_{\bf p}=\nabla_{\bf p} {\cal E}_{\bf p}$.  In an isotropic system ${\bf v}_{\bf p}$ is parallel to ${\bf p}$ and the relation ${\bf v}_{\bf p}={\bf p}/m^\star$ defines the QP effective mass. Thus ${\bf j}_{\bf p}\neq {\bf v}_{\bf p}$, the reason being that due to interactions the moving QP tends to drag part of the electronic medium along with it producing an extra current~[2,3]. 

The QP effective mass is a measurable quantity. The most direct way to determine $m^\star$ would be a measurement of the low-temperature heat capacity $C_V(T)$. It is in fact remarkable~[2,3] that electron-electron interaction effects enter $C_V(T)$ only through $m^\star$: the Landau interaction functions $f_{{\bf p},{\bf p}'}$ are not explicitly invoked~[2,3]. These type of experiments are exceedingly challenging and have not yet been realized with high precision (see Ref.~[38] below and references therein). An alternative tool to access experimentally the QP effective mass (and other Fermi-liquid parameters) is to analyze quantum Shubnikov-de Haas oscillations of the magnetoresistance. Motivated by a large number of recent magnetotransport studies~[20-25] of this type we have revisited~[26] the problem of the microscopic calculation of $m^\star$ 
in a paramagnetic ({\it i.e.} $\zeta=0$) $2D$ EL. Our systematic 
study is based on a generalized $GW$ approximation which makes use of the 
many-body local fields and takes advantage of the 
results of the most recent QMC calculations of the static charge- 
and spin-response of the $2D$ EL~[27]. 
We report extensive calculations for 
the many-body effective mass enhancement $m^\star/m$ over a broad 
range of $r_s$ values. In this respect we critically examine the relative merits of 
the on-shell approximation, commonly used in weak-coupling situations, {\it versus} the actual 
self-consistent solution of the Dyson equation. We show that already for $r_s \simeq 3$ 
and higher, a solution of the Dyson equation proves here necessary in order to obtain 
a well behaved effective mass. Finally we also show that our theoretical 
results for a quasi-$2D$ EL, free of any adjustable fitting parameters, are in good qualitative agreement with some recent measurements in a ${\rm GaAs}/{\rm AlGaAs}$ heterostructure.
\vskip 28 truept

\centerline{\bf 2.  MANY-BODY EFFECTIVE MASS ENHANCEMENT}
\vskip 12 truept
The aim of this Section is to present a brief summary of the theory we have used 
for the retarded QP self-energy $\Sigma_{\rm \scriptstyle ret}({\bf k},\omega)$ of a paramagnetic $2D$ EL 
(that we have summarized in Eqs.~(3) and~(7) below) from which we have calculated $m^\star/m$. The formal justification of Eqs.~(3) and~(7), which essentially rests on both diagrammatic perturbation theory and on the so-called renormalized Hamiltonian approach~[28], can be found in the original works~[29,30], in Ref.~[3], and in Ref.~[26].

To fix the notation we start by introducing the Hamiltonian 
for a $2D$ EL confined to an area $S$,
$$
{\hat {\cal H}}=\sum_{{\bf k},\sigma}\varepsilon_{\bf k}
{\hat c}^{\dagger}_{{\bf k},\sigma}{\hat c}_{{\bf k},\sigma}+\frac{1}{2 S}\sum_{{\bf q}\neq 0}v_{\bf q}\sum_{{\bf k}_1,\sigma_1}\sum_{{\bf k}_2,\sigma_2}{\hat c}^{\dagger}_{{\bf k}_1+{\bf q},\sigma_1}{\hat c}^{\dagger}_{{\bf k}_2-{\bf q},\sigma_2}{\hat c}_{{\bf k}_2,\sigma_2}{\hat c}_{{\bf k}_1,\sigma_1}\,.\eqno(1)
$$
Here ${\hat c}^{\,\dagger}_{{\bf k},\,\sigma}$ and ${\hat c}_{{\bf k},\, \sigma}$ are fermionic creation and annihilation operators which satisfy canonical anticommutation relations, $\varepsilon_{\bf k}=\hbar^2{\bf k}^2/(2m)$ is the single-particle energy, and $v_{\bf q}=2\pi e^2/q$ is the $2D$ Fourier transform of the bare Coulomb interaction $e^2/r$. For later purposes we introduce 
the Fermi wave number $k_F=(2\pi n_{\rm \scriptscriptstyle 2D})^{1/2}=\sqrt{2}/(r_s a_B)$, the Fermi energy $\varepsilon_F=\hbar^2k^2_F/(2m)$ and the quantity $\xi_{\bf k}=\varepsilon_{\bf k}-\varepsilon_F$.

The retarded QP self-energy $\Sigma_{\rm \scriptstyle ret}({\bf k},\omega)$ is written as the sum of two terms,
$$
\Sigma_{\rm \scriptstyle ret}({\bf k},\omega)=
\Sigma_{\rm \scriptstyle SX}({\bf k},\omega)+\Sigma_{\rm \scriptstyle CH}({\bf k},\omega)\eqno(2)
$$
where the first term is called ``screened-exchange" (SX) and the second term is called ``Coulomb-hole" (CH). The frequency $\omega$ is measured from $\varepsilon_F/\hbar$. 

The SX contribution is given by
$$
\Sigma_{\rm \scriptstyle SX}({\bf k},\omega)=
-\int \frac{d^2 {\bf q}}{(2\pi)^2}\,{v_{\bf q} \over \varepsilon({\bf q},\omega-\xi_{{\bf k}+{\bf q}}/\hbar)} \,\Theta(-\xi_{{\bf k}+{\bf q}}/\hbar)\,.\eqno(3)
$$
Here $\Theta(x)$ is the step function and $\varepsilon({\bf q},\omega)$ is a screening function originating 
from effective Kukkonen-Overhauser interactions~[31],
$$
{1 \over \varepsilon({\bf q},\omega)}= 1+v_{\bf q}\,\left[1-G_{+}({\bf q})\right]^2\,\chi_{\rm \scriptstyle C}({\bf q},\omega)+3\,v_{\bf q}\,G^2_{-}({\bf q})\,\chi_{\rm \scriptstyle S}({\bf q},\omega)\,.\eqno(4)
$$
In Eq.~(4) the charge-charge and spin-spin response functions $\chi_{\rm \scriptstyle C}({\bf q}, \omega)$ and $\chi_{\rm \scriptstyle S}({\bf q}, \omega)$ are determined by the spin-symmetric and spin-antisymmetric static~[32] local-field factors $G_{+}({\bf q})$ and $G_{-}({\bf q})$~[27], 
$$
\chi_{\rm \scriptstyle C}({\bf q},\omega)
={\chi_0({\bf q},\omega)\over 1-v_{\bf q}[1-G_+({\bf q})] \chi_0({\bf q},\omega)}\eqno(5)
$$
and 
$$
\chi_{\rm \scriptstyle S}({\bf q},\omega)
={\chi_0({\bf q},\omega) \over 1+v_{\bf q}G_{-}({\bf q})\,\chi_0({\bf q},\omega)}\,,\eqno(6)
$$
$\chi_0({\bf q},\omega)$ being the Lindhard-Stern response function of a noninteracting $2D$ electron gas~[34]. 
Note that $\Sigma_{\rm \scriptstyle SX}({\bf k},\omega)$ is just an ordinary exchange-like self-energy built from the Kukkonen-Overhauser effective interactions instead of bare Coulomb interactions, which would lead to the frequency-independent Hartree-Fock self-energy first calculated for the $2D$ EL by Chaplik~[35].

The CH contribution to the retarded self-energy is given by
$$
\Sigma_{\rm \scriptstyle CH}({\bf k},\omega)=
-\int {d^2 {\bf q} \over (2\pi)^2}
\,v_{\bf q}\int_{0}^{+\infty}\,{d \Omega \over \pi}\,\,{\Im m[\varepsilon^{-1}({\bf q},\Omega)]\over 
\omega-\xi_{{\bf k}+{\bf q}}/\hbar-\Omega+i\delta}\,,\eqno(7)
$$
where $\delta$ is a positive infinitesimal. The real part of the retarded self-energy is readily obtained 
from Eqs.~(3) and~(7) with the result
$$
\eqalignno{
\Re e\Sigma_{\rm \scriptstyle ret}({\bf k},\omega)&=-\int {d^2 {\bf q} \over (2\pi)^2}\,
v_{\bf q}\Re e[\varepsilon^{-1}({\bf q},\omega-\xi_{{\bf k}+{\bf q}}/\hbar)]
\,\Theta(-\xi_{{\bf k}+{\bf q}}/\hbar)\cr
& \quad -\int {d^2 {\bf q} \over (2\pi)^2}\,v_{\bf q}{\cal P}\int_{0}^{+\infty}\,{d \Omega \over \pi}\,
{\Im m[\varepsilon^{-1}({\bf q},\Omega)] \over \omega-\xi_{{\bf k}+{\bf q}}/\hbar-\Omega}\,.&(8)
\cr}
$$
Once the real part of the QP self-energy is known, 
the QP excitation energy $\delta {\cal E}_{\rm QP}({\bf k})$, which is the QP energy measured from the 
chemical potential $\mu$ of the interacting EL, can be calculated 
by solving self-consistently the Dyson equation
$$
\delta {\cal E}_{\rm QP}({\bf k})=\xi_{\bf k}+\left. 
\Re e \Sigma^{\rm R}_{\rm \scriptstyle ret}({\bf k},\omega)\right|_
{\omega=\,\delta {\cal E}_{\rm QP}({\bf k})/\hbar}\,,\eqno(9)
$$
where $\Re e\Sigma^{\rm R}_{\rm \scriptstyle ret}({\bf k},\omega)=
\Re e \Sigma_{\rm \scriptstyle ret}({\bf k},\omega)-\Sigma_{\rm \scriptstyle ret}(k_F,0)$. For later purposes we introduce at this point the so-called on-shell approximation (OSA). This amounts to approximating the QP excitation energy by calculating $\Re e \Sigma^{\rm R}_{\rm \scriptstyle ret}({\bf k},\omega)$ in Eq.~(9) at the frequency $\omega=\xi_{\bf k}/\hbar$ corresponding to the single-particle energy, that is
$$
\delta {\cal E}_{\rm QP}({\bf k})\simeq \xi_{\bf k}+\left. \Re e \Sigma^{\rm R}_{\rm \scriptstyle ret}({\bf k},\omega)\right|_{\omega=\xi_{\bf k}/\hbar}\,.\eqno(10)
$$

The effective mass $m^\star$ can be calculated from the QP excitation energy by means of the relationship
$$
\frac{1}{m^\star}=\frac{1}{\hbar^2 k_F} \left. {d \delta {\cal E}_{\rm QP}(k) \over dk}\right|_{k=k_F}\,.\eqno(11)
$$

As remarked above, $\delta {\cal E}_{\rm QP}(k)$ may be calculated either by 
solving self-consistently the Dyson equation~(9) or by using the OSA in Eq.~(10). 
In what follows the identity
$$
{d \Re e\Sigma^{\rm R}_{\rm \scriptstyle ret}(k,\omega(k)) \over dk}=\left.\partial_k \Re e\Sigma^{\rm R}_{\rm \scriptstyle ret}(k,\omega)\right|_{\omega=\omega(k)}+\left.
\partial_{\omega} \Re e\Sigma^{\rm R}_{\rm \scriptstyle ret}(k,\omega)\right|_{\omega=\omega(k)}
{d\omega(k) \over dk}\eqno(12)
$$
will be used, $\omega(k)$ being an arbitrary function of $k$. 

Using Eqs.~(11) and~(12) with 
$\omega(k)=\delta {\cal E}_{\rm QP}(k)/\hbar$ we find that the effective mass 
$m^\star_{\rm D}$ calculated within the Dyson scheme is given by
$$
{m^*_{\rm D}\over m}={Z^{-1} \over \displaystyle 1+(m/\hbar^{2} k_F)\left.
\partial_k \Re e \Sigma^{\rm R}_{\rm \scriptstyle ret}(k,\omega)\right|_{k=k_F,\omega=0}}\,.\eqno(13)
$$
The renormalization constant $Z$ that 
measures the discontinuity of the momentum distribution at $k=k_F$ is given by
$$
Z={1 \over 1-\hbar^{-1} \left.\partial_{\omega} \Re e \Sigma^{\rm R}_{\rm \scriptstyle ret}(k,\omega)\right|_{k=k_F,\omega=0}}\,.\eqno(14)
$$
The normal Fermi-liquid assumption, $0<Z\leq 1$, implies $\left.\partial_{\omega} \Re e \Sigma^{\rm R}_{\rm \scriptstyle ret}(k,\omega)\right|_{k=k_F,\omega=0} \leq 0$. 
Thus we see that the effective mass $m^\star_{\rm D}$ can diverge at a finite 
value of $r_s$ by one of two mechanisms: (i) the partial derivative 
of $\Sigma^{\rm R}_{\rm \scriptstyle ret}$ with respect to $\omega$,  
$\left.\partial_{\omega} \Re e \Sigma^{\rm R}_{\rm \scriptstyle ret}(k,\omega)\right|_{k=k_F,\omega=0}$ 
going to minus infinity at some finite value of $r_s$~[36]; (ii) the partial derivative of $\Sigma^{\rm R}_{\rm \scriptstyle ret}$ with respect to $k$, $\left.\partial_k \Re e \Sigma^{\rm R}_{\rm \scriptstyle ret}(k,\omega)\right|_{k=k_F,\omega=0}$ going to $-\hbar^2 k_F/m$ at some finite value of $r_s$. 

Neither possibility is 
realized in our calculation:  the first is barred {\it a priori} by the 
fact that the analytic expression for the frequency derivative of 
$\Sigma^{\rm R}_{\rm \scriptstyle ret}$ is always finite at finite $r_s$;  the second is found {\it 
a posteriori} not to occur since the momentum derivative of $\Sigma^{\rm R}_{\rm \scriptstyle ret}$ is 
positive up to the largest $r_s$ considered (see below).

On the other hand, using Eqs.~(11) and~(12) with 
$\omega(k)=\xi_{\bf k}/\hbar$ we find that the effective mass $m^\star_{\rm OSA}$ within the OSA is given by
$$
{m^\star_{\rm OSA} \over m}={1 \over 1+(m/\hbar^{2} k_F)\left.\partial_k \Re e \Sigma^{\rm R}_{\rm \scriptstyle ret}(k,\omega)\right|_{k=k_F,\omega=0}+(1-Z^{-1})}\,.\eqno(15)
$$
Of course, Eq.~(15) is a valid approximation to the effective mass 
in the weak coupling limit, as can be seen by expanding Eq.~(13) for 
small values of $\Sigma^{\rm R}_{\rm \scriptstyle ret}$:  however its application becomes 
problematic at large values of $r_s$.  In particular, we see that 
because $Z$ decreases monotonically with 
increasing $r_s$, there must necessarily be a critical value of $r_s$ 
for which the denominator of Eq.~(15) vanishes and $m^\star_{\rm OSA}$ diverges. 
A recent paper by Zhang and Das Sarma~[37] 
infers from this fact a true divergence of the effective mass within the RPA. 
In our view, however, this must be considered an artifact of Eq.~(15). 
The unphysical character of the behavior $m^\star_{\rm OSA}\rightarrow \infty$ is revealed by the fact that the 
divergence is driven by a negative but finite value of 
$\partial_{\omega} \Re e \Sigma^{\rm R}_{\rm \scriptstyle ret}(k_F,0)$, 
whereas we know, from the general analysis given above, 
that a genuine divergence would have to be driven either by an 
infinite $\partial_{\omega} \Re e \Sigma^{\rm R}_{\rm \scriptstyle ret}(k_F,0)$ 
or by a  negative $\partial_k \Re e \Sigma^{\rm R}_{\rm \scriptstyle ret}(k_F,0)$ 
becoming equal to $-\hbar^2k_F/m$. 
We conclude that there is no evidence, within the present theory, for a physically relevant 
divergence of the effective mass.
\vskip 28 truept

\centerline{\bf 3.  NUMERICAL RESULTS}
\vskip 12 truept
We turn now to a presentation of our main numerical results for $m^\star/m$. 
In all figures the labels ``RPA", ``$G_+$" and ``$G_+ \& G_-$" refer to three possible choices for the local-field factors: ``RPA" refers to the case in which local-field factors are not included, ``$G_+$" to the case in which the antisymmetric spin-spin local field is set to zero ({\it i.e.} spin-density fluctuations are not allowed), and finally ``$G_+ \& G_-$" refers to the full theory including both charge- and spin-density fluctuations.

In Fig.~1 we show our numerical results for $m^\star_{\rm D}$ and $m^\star_{\rm OSA}$. The effective mass enhancement is substantially smaller in the Dyson-equation calculation than in the OSA, the reason being that a large cancellation occurs between numerator and denominator in Eq.~(13). In both calculations 
the combined effect of charge and spin fluctuations is to enhance the effective mass over the RPA result, whereas the opposite effect is found if only charge fluctuations are included -- a manifestly incorrect result that neglects the spinorial nature of the electron. For completeness we have also included in Fig.~1 
the variational QMC results of Kwon {\it et al.}~[11]. The reader should bear in mind that the effective mass is not a ground-state property and thus its evaluation by the QMC technique is quite delicate, as it involves the construction of excited states. There clearly is quantitative disagreement between our ``best" theoretical results 
(the ``$G_+\& G_-/{\rm D}$" predictions) and the QMC data.

\topinsert
\input psfig.sty
\centerline{\hskip-12mm\psfig{figure=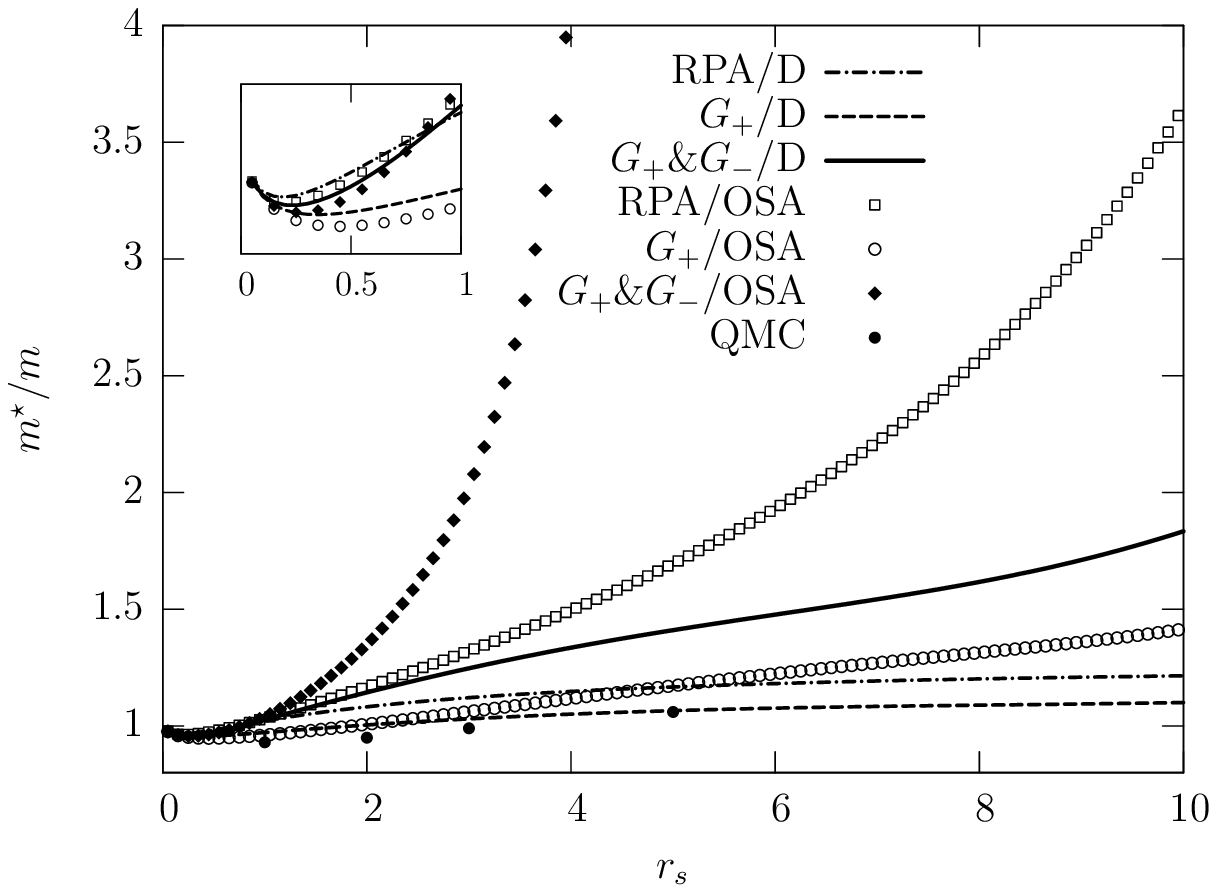,height=9truecm,width=12truecm,angle=0}}
\noindent
{\bf Figure 1.} 
Effective mass enhancement as a function of $r_s$ for $0 \leq r_s \leq 10$. The inset shows 
an enlargement of the results for $r_s \leq 1$. The lines show the results from Eq.~(13), 
while the symbols (except for the dots) are from Eq.~(15). The QMC data (dots) are from Ref.~[11].\vskip 12truept
\endinsert

In Fig.~2 we show the behavior of the two terms in the denominator of Eq.~(15) 
as functions of $r_s$. This figure clearly shows how a spurious divergence can arise in $m^\star_{\rm OSA}$: 
for instance, within the RPA the denominator in Eq.~(15) has a zero at 
$r_s\simeq 15.5$ (see the inset in Fig.~2). Our numerical evidence, within the three theories 
we have studied, is that indeed (i) $\partial_\omega \Re e \Sigma^{\rm R}_{\rm \scriptstyle ret}(k_F,0)$ 
is negative as it should for a normal Fermi-liquid, and monotonically increasing in absolute value 
as a function of $r_s$; and (ii) $\partial_k \Re e \Sigma^{\rm R}_{\rm \scriptstyle ret}(k_F,0)$ 
is positive and monotonically increasing too. Within the theory outlined in Ref.~[26], which 
uses as a key ingredient the Kukkonen-Overhauser effective screening function in Eq.~(4), 
the effect of a charge-only local field is to shift this divergence to higher values of $r_s$, 
while the opposite occurs upon including both charge and spin fluctuations. For instance, within 
the ``$G_+ \& G_-/{\rm OSA}$" theory the divergence occurs near $r_s=5$.
\topinsert
\input psfig.sty
\centerline{\hskip-12mm\psfig{figure=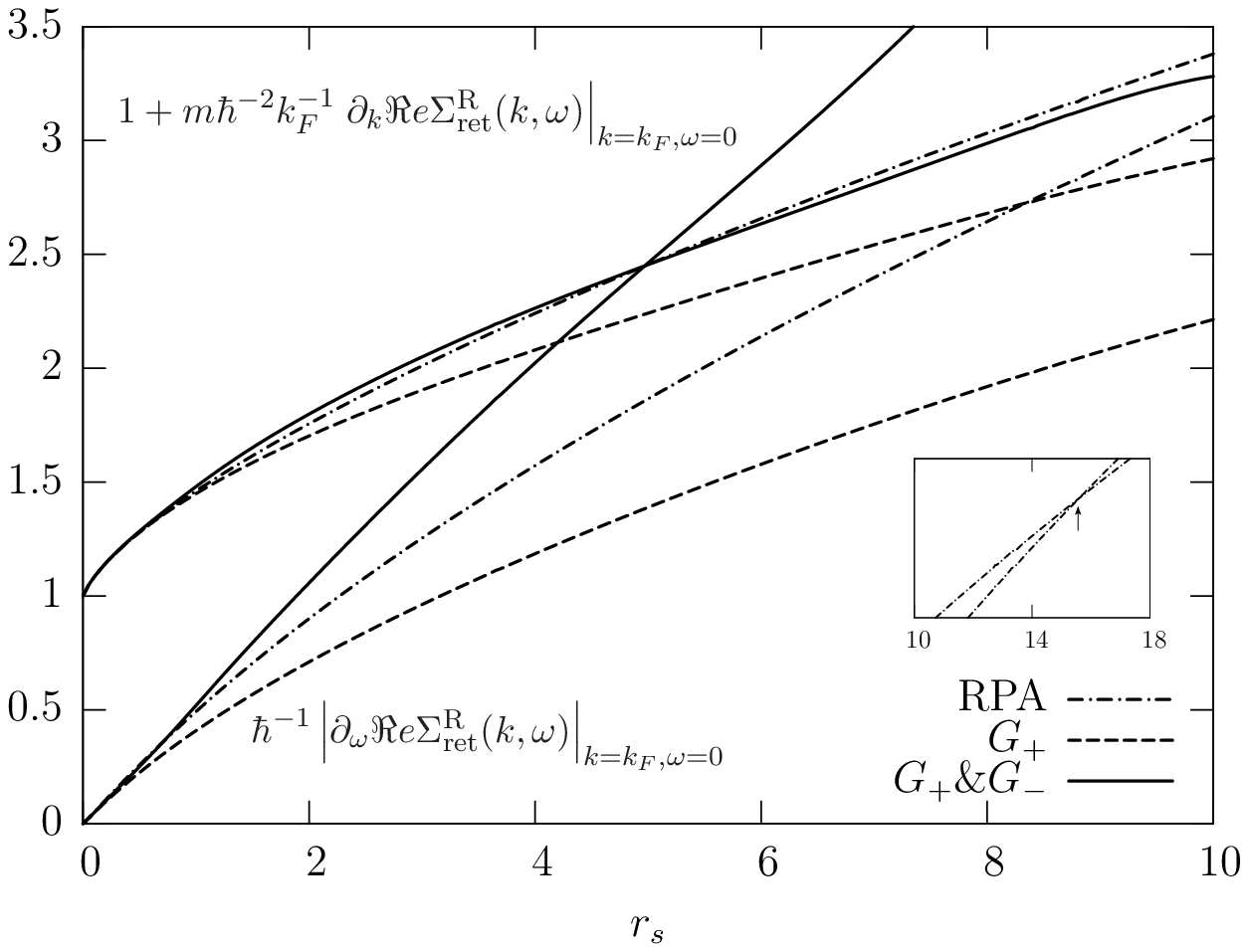,height=9truecm,width=12truecm,angle=0}}
\noindent
{\bf Figure 2.}
Illustrating the divergence of the effective mass within the OSA. The 
three curves starting from unity at $r_s=0$ refer to the quantity 
$1+m\hbar^{-2}k^{-1}_F\partial_k \Re e \Sigma^{\rm R}_{\rm \scriptstyle ret}(k_F,0)$, and 
the other three curves to $\hbar^{-1}|\partial_{\omega} \Re e \Sigma^{\rm R}_{\rm \scriptstyle ret}(k_F,0)|=Z^{-1}-1$. 
The intersection of two lines with the same line-style in the two sets of curves corresponds to a zero 
in the denominator of Eq.~(15) and thus to a divergence in $m^\star_{\rm OSA}$. 
The inset shows this divergence occurring within the RPA at $r_s\simeq 15.5$.\vskip 12truept
\endinsert

We now pass to illustrate how our theory compares with experiments. 
A full analysis of the published data for the effective mass of carriers in 
Si-MOSFET's~[20,21] would require 
a more complete theoretical study, mainly to account for the two-valley nature of the material. 
We will focus here instead on the experimental results of Tan {\it et al.}~[38] in a 
${\rm GaAs}/{\rm AlGaAs}$ heterojunction-insulated gate field-effect transistor (HIGFET) of 
exceedingly high quality. A quantitative comparison between theory and experiment 
would also require a refined treatment of a 
series of effects such as those due to (i) the detailed band-structure of the host semiconductor, (ii) disorder, and (iii) finite temperature~[39]. Note also that the Shubnikov-de Haas measurements 
of Tan {\it et al.}~[38] are performed in a small but obviously {\it finite} magnetic field $B$ and, in general, 
the ground-state of an EL in a small finite $B$ field~[40] can be profoundly different from that at $B=0$. In our zero-field calculations we have: (i) included band-structure effects only through the ${\rm GaAs}$ band mass $m_{\rm b}=0.067\,m$; (ii) neglected possible effects due to disorder because the concentration of background impurities $n_i$ in the HIGFET used in Ref.~[38] has been estimated~[41] to be $n_i \approx 5\times 10^{12}\,{\rm cm}^{-3}$ which is a very low number; and (iii) neglected thermal effects even though in Ref.~[38] the temperature of the dilution refrigerator was kept relatively high ($100 \lesssim T \lesssim 400\,{\rm mK}$) to avoid the quantum Hall regime in which the Shubnikov-de Haas oscillations become non-sinusoidal. A simple inspection shows however that, assuming that for the highest temperature $T=400\,{\rm mK}$ the EL is in thermal equilibrium with the refrigerator, the ratio of thermal to Fermi energy is quite small even at the lowest densities, {\it e.g.} $k_BT/\varepsilon_F\approx 0.03$ for $r_s=6$.

We thus restrict our analysis solely 
to the effect of finite sample thickness, by discussing how a softened Coulomb potential modifies 
$m^\star$ against the strictly-$2D$ results shown in Fig.~1. 
The expectation is that the QP effective mass will be noticeably smaller 
when a softened Coulomb interaction is at work.

We have thus recalculated $m^\star$ after renormalizing the bare Coulomb potential by means of a form factor to take into account the finite width of the EL in the HIGFET used in Ref.~[38]. The appropriate renormalized potential is given by $V_{\bf q}=v_{\bf q}{\cal F}(qd)/{\bar \kappa}$, where 
$$
{\cal F}(x)=\left(1+{\kappa_{\rm ins} \over \kappa_{\rm sc}}\right){8+9x+3x^2 \over 16(1+x)^3}+\left(1-{\kappa_{\rm ins} \over \kappa_{\rm sc}}\right){1 \over 2(1+x)^6}\,,\eqno(16)
$$
with $d=[\hbar^2 \kappa_{\rm sc}/(48\pi m_{\rm b} e^2n^\star)]^{1/3}$ representing an effective width of the quasi-$2D$ EL~[42]. Here $\kappa_{\rm ins}=10.9$ and $\kappa_{\rm sc}=12.9$ are the dielectric constants of 
the insulator and of the space charge layer, ${\bar \kappa}$ is their average and $n^\star=n_{\rm depl}+11n_{\rm \scriptscriptstyle 2D}/32$, the depletion layer charge density $n_{\rm depl}$ being essentially zero (see Ref.~[18] of Ref.~[41]) in the experiments of Ref.~[38]. Note that the renormalized potential does not contain any adjustable fitting parameter. The results that we obtain with the softened potential $V_{\bf q}$ are shown in Fig.~3 against the experimental results of Tan {\it et al.}~[38]. 
A {\it caveat} to keep in mind is that we have used the same local-field factors as for a zero-thickness $2D$ EL~[27]
in the lack of a better choice. Thus the results labeled by ``$G_+$" and ``$G_+\&G_-$" in Fig.~3 contain the effect of finite thickness only through the renormalization of the Coulomb potential. We believe that the explicit dependence of the local fields on the finite width of the $2D$ EL should not change the results of Fig.~3 in a substantial manner. 

As it appears from Fig.~3, a quite satisfactory qualitative agreement exists between theory and experiment, considering the oversimplified model we have used for the quasi-$2D$ EL in the heterojunction of Ref.~[38]. Nevertheless, in detail the discrepancies are quite substantial. In particular both the RPA and ``$G_+\&G_-$"  theories, which give rather similar results, underestimate the considerable effective mass enhancement at strong-coupling, {\it i.e.} for $r_s \gtrsim 5$. A comparison in the weak-coupling regime $r_s \lesssim 1$ would be very helpful but unfortunately is not possible due to the lackness of experimental data. 
\topinsert
\input psfig.sty
\centerline{\hskip-12mm\psfig{figure=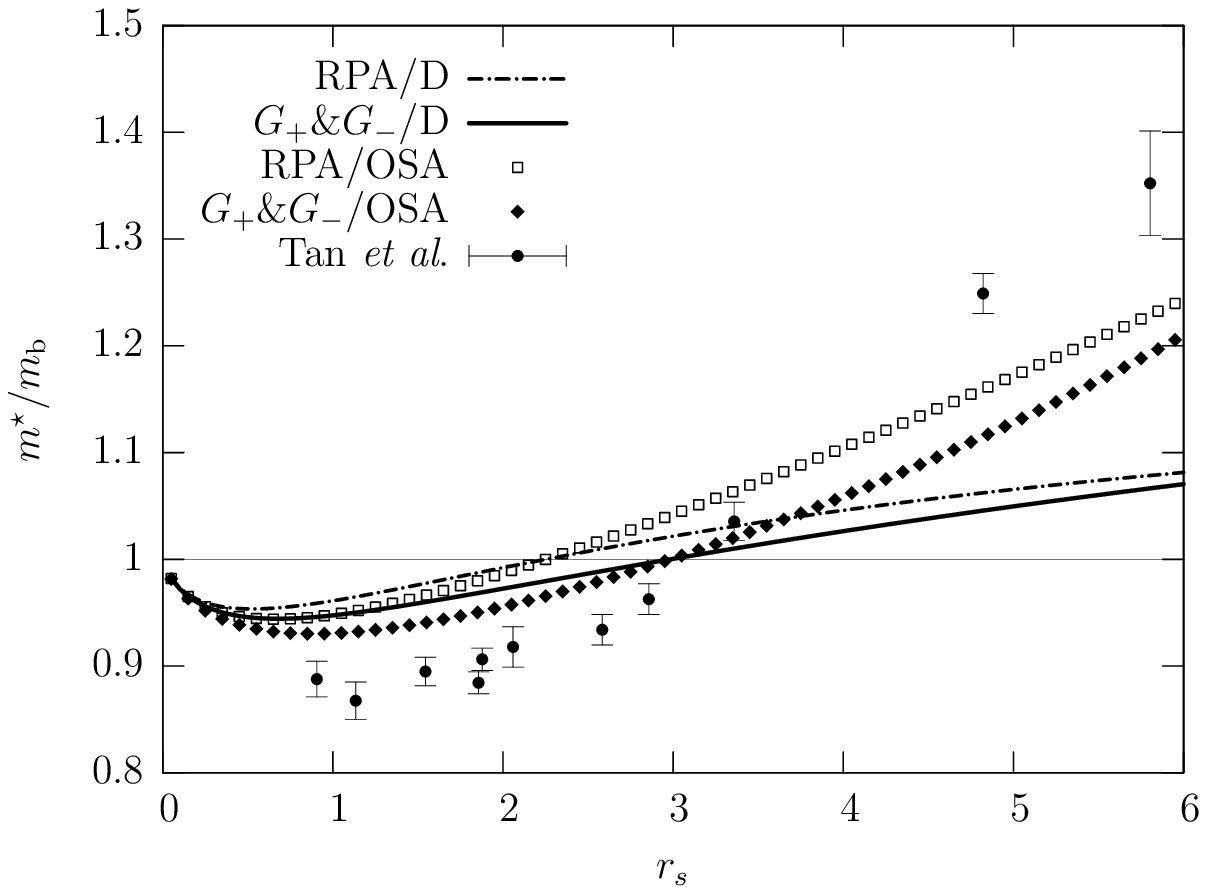,height=9truecm,width=12truecm,angle=0}}
\noindent
{\bf Figure 3.}
Effective mass enhancement for a $2D$ EL confined in a ${\rm GaAs}/{\rm AlGaAs}$ triangular quantum well of the type used in Refs.~[24] and~[38]. The theoretical results are shown in the same notation of Fig.~1. 
The experimental results (Tan {\it et al.}) are from Ref.~[38].\vskip 12truept
\endinsert

\vskip 28 truept
\centerline{\bf 4.  CONCLUSIONS}
\vskip 12 truept
In summary, we have revisited the problem of the microscopic
calculation of the many-body effective mass 
enhancement in a $2D$ EL. We have performed a systematic study 
based on the many-body local-fields theory, taking advantage of the results of 
the most recent QMC calculations of the static charge and spin response of the EL 
expressed through static local-field factors. We have presented results for the effective mass 
enhancement over a wide range of electron densities. 
In this respect we have critically examined the merits 
of the OSA {\it versus} the Dyson-equation calculation. Depending on 
the local-field factors, the OSA predicts a spurious divergence of the effective mass at 
strong coupling and a solution of the Dyson equation is therefore necessary in order 
to obtain the correct value of the effective mass within Fermi-liquid theory. The comparison 
with the experimental data of Ref.~[38] shows good qualitative agreement but substantial discrepancies especially at strong coupling calling for further theoretical work and computer simulations.
\vskip 28 truept

\centerline{\bf ACKNOWLEDGMENTS}
\vskip 12 truept
We are grateful to George Simion for useful discussions and to Yanwen Tan and Jun Zhu for sharing with us their considerable physical insight and their experimental results. One of us (M.P.) gratefully acknowledges the hospitality of Giovanni Vignale at the Department of Physics and Astronomy of the University of Missouri-Columbia and 
partial support from Washington University and from the Division of Undergraduate 
Education of the U.S. National Science Foundation through grant 
DUE-0127488 to the Shodor Education Foundation, Inc.
This work was partially supported by MIUR through the PRIN2001 and PRIN2003 programs. 
G.V. acknowledges support from NSF Grant No. DMR 0313681.
\vskip 28 truept

\centerline{\bf REFERENCES}
\vskip 12 truept

\item{[1]}
D.~M.~Ceperley, {\it Nature} {\bf 397}, 386 (1999).
\item{[2]}
D.~Pines and P.~Nozi\'eres, {\it The Theory of Quantum Liquids} 
(W.~A.~Benjamin, Inc., New York, 1966).
\item{[3]}
G.~F.~Giuliani and G.~Vignale, {\it Quantum Theory of the Electron Liquid} 
(Cambridge University Press, Cambridge, 2005).
\item{[4]}
K.~S.~Singwi and M.~P.~Tosi, in {\it Solid State
Physics} {\bf 36}, ed. H.~Ehrenreich, F.~Seitz and D.~Turnbull
(Academic, New York, 1981), p. 177.
\item{[5]}
A.~W.~Overhauser, {\it Phys.~Rev.~B} {\bf 3}, 1888 (1971).
\item{[6]}
N.~H.~March, {\it Electron Correlation in the Solid State} (Imperial College Press, London, 1999).
\item{[7]} 
E.~P.~Wigner, {\it Phys.~Rev.} {\bf 46}, 1002 (1934).
\item{[8]}
D.~M.~Ceperley and B.~J.~Alder, {\it Phys.~Rev.~Lett.} {\bf 45}, 566 (1980); 
B.~J.~Alder, D.~M.~Ceperley, and E.~L.~Pollock, {\it Int.~J.~Quant.~Chem.} {\bf 16}, 49 (1982).
\item{[9]}
B.~Tanatar and D.~M.~Ceperley, {\it Phys.~Rev.~B} {\bf 39}, 5005 (1989). 
\item{[10]}
S.~Moroni, D.~M.~Ceperley, and G.~Senatore, {\it Phys.~Rev.~Lett.} {\bf 69}, 1837 (1992). 
\item{[11]}
Y.~Kwon, D.~M.~Ceperley, and R.~M.~Martin, {\it Phys.~Rev.~B} {\bf 50}, 1684 (1994).
\item{[12]}
S.~Moroni, D.~M.~Ceperley and G.~Senatore, {\it Phys.~Rev.~Lett.} {\bf 75}, 689 (1995).
\item{[13]}
F.~Rapisarda and G.~Senatore, {\it Austr.~J.~Phys.} {\bf 49}, 161 (1996).
\item{[14]}
G.~Ortiz, M.~Harris, and P.~Ballone, {\it Phys.~Rev.~Lett.} {\bf 82}, 5317 (1999).
\item{[15]}
G.~Senatore, S.~Moroni, and D.~M.~Ceperley, in {\it Quantum Monte Carlo Methods in Physics and
Chemistry}, ed. M.~P.~Nightingale and C.~J.~Umrigar (Kluwer, Dordrecht, 1999).
\item{[16]}
D.~Varsano, S.~Moroni, and G.~Senatore, {\it Europhys.~Lett.} {\bf 53}, 348 (2001).
\item{[17]}
F.~H.~Zong, C.~Lin, and D.~M.~Ceperley, {\it Phys.~Rev.~E} {\bf 66}, 036703 (2002).
\item{[18]}
C.~Attaccalite, S.~Moroni, P.~Gori-Giorgi, and G.~Bachelet, {\it Phys.~Rev.~Lett.} {\bf 88}, 256601 (2002).
\item{[19]} 
L.~D.~Landau, {\it Sov.~Phys.~JEPT} {\bf 3}, 920 (1957).
\item{[20]}
A.~A.~Shashkin, S.~V.~Kravchenko, V.~T.~Dolgopolov, and T.~M.~Klapwijk, {\it Phys.~Rev.~Lett.} 
{\bf 87}, 086801 (2001) and {\it Phys.~Rev.~B} {\bf 66}, 073303 (2002).
\item{[21]}
V.~M.~Pudalov, M.~E.~Gershenson, H.~Kojima, N.~Butch, E.~M.~Dizhur, G.~Brunthaler, A.~Prinz, and G.~Bauer, {\it Phys.~Rev.~Lett.} {\bf 88}, 196404 (2002).
\item{[22]}
E.~Tutuc, S.~Melinte, and M.~Shayegan, {\it Phys.~Rev.~Lett.} {\bf 88}, 036805 (2002).
\item{[23]}
H.~Noh, M.~P.~Lilly, D.~C.~Tsui, J.~A.~Simmons, E.~H.~Hwang, S.~Das~Sarma, L.~N.~Pfeiffer, and K.~W.~West, 
{\it Phys.~Rev.~B} {\bf 68}, 165308 (2003).
\item{[24]}
J.~Zhu, H.~L.~Stormer, L.~N.~Pfeiffer, K.~W.~Baldwin, and K.~W.~West, {\it Phys.~Rev.}\break 
{\it Lett.} {\bf 90}, 056805 (2003).
\item{[25]}
K.~Vakili, Y.~P.~Shkolnikov, E.~Tutuc, E.~P.~De~Poortere, and M.~Shayegan, {\it Phys.~Rev.~Lett.} {\bf 92}, 226401 (2004).
\item{[26]}
R.~Asgari, B.~Davoudi, M.~Polini, G.~F.~Giuliani, M.~P.~Tosi, 
and G.~Vignale, {\it Phys. Rev. B} in press and cond-mat/0406676.
\item{[27]}
B.~Davoudi, M.~Polini, G.~F.~Giuliani, and M.~P.~Tosi, {\it Phys.~Rev.~B} {\bf 64}, 153101 and 233110 (2001).
\item{[28]}
D.~R.~Hamann and A.~W.~Overhauser, {\it Phys.~Rev.} {\bf 143}, 183 (1966).
\item{[29]}
G.~Vignale and K.~S.~Singwi, {\it Phys.~Rev.~B} {\bf 32}, 2156 (1985); 
T.~K.~Ng and K.~S.~Singwi, {\it ibid.} {\bf 34}, 7738 and 7743 (1986).
\item{[30]}
S.~Yarlagadda and G.~F.~Giuliani, {\it Solid~State~Commun.} {\bf 69}, 677 (1989); 
G.~E.\break Santoro and G.~F.~Giuliani, {\it Phys.~Rev.~B} {\bf 39}, 12818 (1989);
S.~Yarlagadda and G.~F.~Giuliani, {\it ibid.} {\bf 40}, 5432 (1989); 
S.~Yarlagadda and G.~F.~Giuliani, {\it ibid.} {\bf 49}, 7887 and 14188 (1994).
\item{[31]}
C.~A.~Kukkonen and A.~W.~Overhauser, {\it Phys.~Rev.~B} {\bf 20}, 550 (1979).
\item{[32]}
Although the local-field factors are frequency-dependent quantities, we have made 
the common, and to a certain extent uncontrolled, approximation of neglecting their frequency 
dependence. Recent studies~[33] have explored such a dependence in the long-wavelength 
limit ${\bf q}\rightarrow 0$, but clearly the knowledge of the full dependence on wave number is 
necessary for correctly carrying out the type of calculations that we are interested in this work. 
A possible role of dynamical exchange-correlation effects is currently under study. 
\item{[33]}
H.~M.~B\"ohm, S.~Conti, and M.~P.~Tosi, {\it J.~Phys.:~Condens.~Matter} {\bf 8}, 781 (1996);
S.~Conti, R.~Nifos\`\i, and M.~P.~Tosi, {\it ibid.} {\bf 9}, L475 (1997);	
R.~Nifos\`\i, S.~Conti, and M.~P.~Tosi, {\it Phys.~Rev.~B} {\bf 58}, 12758 (1998); 
G.~Vignale and W.~Kohn, in {\it Electronic Density Functional Theory, Recent Progress and New Directions}, 
ed. J.~Dobson, G.~Vignale, and M.~P.~Das (Plenum, New York, 1998) p.~199; 
Z.~Qian and G.~Vignale, {\it Phys.~Rev.~B} {\bf 65}, 235121 (2002) and {\bf 68}, 195113 (2003).	
\item{[34]}
F.~Stern, {\it Phys.~Rev.~Lett.} {\bf 18}, 546 (1967).
\item{[35]}
A.~V.~Chaplik, {\it Sov.~Phys.~JETP} {\bf 33}, 997 (1971).
\item{[36]}
In this case the normal Fermi-liquid assumption breaks down and a singular behavior of 
$m^\star_{\rm D}$ could be interpreted as a quantum phase transition of the $2D$ EL 
to a non-Fermi-liquid state.
\item{[37]}
Y.~Zhang and S.~Das Sarma, cond-mat/0312565.
\item{[38]}
Y.~-W.~Tan, J.~Zhu, H.~L.~Stormer, L.~N.~Pfeiffer, K.~W.~Baldwin, and K.~W.\break West, {\it Phys.~Rev.~Lett.} in press and cond-mat/0412260.
\item{[39]}
S.~Das~Sarma, V.~M.~Galitski, and Y.~Zhang, {\it Phys.~Rev.~B} {\bf 69}, 125334 (2004).
\item{[40]}
A.~A.~Koulakov, M.~M.~Fogler, and B.~I.~Shklovskii,
{\it Phys.~Rev.~Lett.} {\bf 76}, 499 (1996); R.~Moessner
and J.~T.~Chalker, {\it Phys.~Rev.~B} {\bf 54}, 5006 (1996); 
M.~P.~Lilly, K.~B.~Cooper, J.~P.~Eisenstein, L.~N.~Pfeiffer, and
K.~W.~West, {\it Phys.~Rev.~Lett.} {\bf 82}, 394 (1999); 
R.~R.~Du, D.~C.~Tsui, H.~L.~Stormer, L.~N.~Pfeiffer, K.~W.~Baldwin, and
K.~W.~West, {\it Solid~State~Commun.} {\bf 109}, 389 (1999); K.~B.~Cooper,
M.~P.~Lilly, J.~P.~Eisenstein, L.~N.~Pfeiffer, and
K.~W.~West, {\it Phys.~Rev.~B} {\bf 60}, R11285 (1999);
F.~D.~M.~Haldane, E.~H.~Rezayi, and K.\break Yang, {\it Phys.~Rev. Lett.} {\bf 85}, 5396 (2000).
\item{[41]}
S.~De~Palo, M.~Botti, S.~Moroni, and G.~Senatore, cond-mat/0410145.
\item{[42]}
T.~Ando, A.~B.~Fowler, and F.~Stern, {\it Rev.~Mod.~Phys.} {\bf 54}, 437 (1982).
\end{document}